\documentstyle[preprint,amsfonts,aps]{revtex}
\tightenlines
\newcommand{\rv}{{\bf r}}

\newcommand{\Rv}{{\bf R}}

\newcommand{\Mvec}{{\bf M}}

\newcommand{\rd}{r_{12}}
\newcommand{\half}{{1\over 2}}

\newcommand\eqref[1]{Eq.~(\ref{#1})}

\newcommand\cref[1]{Ref.~\cite{#1}}
\newcommand\abbref[1]{Fig.~\ref{#1}}

\newcommand\abbpref[1]{Fig.~\protect\ref{#1}}

\newcommand{\nn}{\nonumber}
\newcommand\av[1]{\left\langle #1 \right\rangle}
\newcommand\ts{T^\ast}
\newcommand\ms{{m^\ast}}
\newcommand\rhos{\rho^\ast}

\begin{document}
\draft
\title{Bulk and Surface Properties of Dipolar Fluids}
\author{B. Groh and S. Dietrich}
\address{
  Fachbereich Physik, Bergische Universit\"at Wuppertal, \\
 D--42097 Wuppertal, Federal Republic of Germany}
\date{\today}

\maketitle


\begin{abstract}

Based on density-functional theory we analyze the full phase diagram,
the occurrence of long-ranged orientational order, and the structural
properties of dipolar fluids.
As a model system we consider the
Stockmayer fluid that consists of spherical particles interacting via
a Lennard-Jones potential plus dipolar forces. For sufficiently
strong dipole moments one finds  a region where a fluid phase
with long-ranged orientational order is stable. For all sample shapes
with the exception of a long thin needle this phase exhibits a
spatially inhomogeneous magnetization which depends on the actual
shape. We determine the details of the magnetization structure
in a cubic sample in the absence and in the presence of an external
magnetic field. One obtains a vortexlike structure with an escape of the
magnetization into the axis  direction near the vortex axis and two
point defects where the absolute value of the magnetization is
strongly reduced.
If the spherical cores of the particles are replaced by elongated or
oblate shapes a nematic phase without spontaneous magnetization is
also possible due to the anisotropic steric interactions. We study the
interplay of this nematic ordering with ferromagnetism in fluids of dipolar
hard ellipsoids.
Orientational order arises locally in the isotropic fluid phases near
the liquid-gas interface of the Stockmayer fluid. Density-functional
theory allows us to determine density and orientational order
profiles as well as the surface tension of this interface.

\end{abstract}

\newpage

\section{Introduction}

Spontaneous symmetry breaking induced by phase transitions is one
of the central themes of statistical physics. From a
homogeneous and isotropic liquid phase transitions can occur to phases
with positional or orientational long-ranged order of the
individual particles. Here we are mainly interested in the case of
pure orientational order. Well known examples of such systems are the nematic liquid
crystals. Initially, it was suspected that dipolar interactions
are responsible for the formation of the nematic state \cite{Born}.
Later, it became clear that the short-ranged steric interactions
between strongly anisotropic (elongated or disklike) nonpolar
molecules are sufficient to generate a nematic phase. On the other
hand, it is an interesting question whether dipolar interactions
alone, i.e., in the absence of anisotropic steric interactions,
are also capable of stabilizing an orientationally ordered liquid.
That this is indeed possible has been shown for the first time some
years ago in computer simulations of a dipolar soft sphere fluid by
Wei and Patey \cite{Patey1,Patey2}. Similar observations were made for
dipolar hard spheres \cite{Weis1,Weis2,Grest2} and the Stockmayer
fluid \cite{Grest}, in which the spherically symmetric Lennard-Jones
potential is added to the dipolar interactions.  Depending on whether
one deals with electric or magnetic dipole moments this new phase, in
which the dipoles are aligned  over macroscopic distances, is
called a ferroelectric or ferromagnetic liquid. In the following we
will adopt the magnetic language keeping in mind that all results are
equally valid in the electric case. Theoretically this phase
transition was predicted within mean-field theory for a dipolar
lattice gas \cite{Sano} and by a generalized van der Waals theory
\cite{WZ}. A density-functional theory using as input  direct correlation
functions of the isotropic liquid obtained from an integral equation
theory qualitatively confirmed these findings
\cite{Patey3}. However, more recent work along the same lines
\cite{Klapp1} predicts that the ferromagnetic liquid is
metastable with respect to freezing. It might be possible to improve
upon this approach by directly calculating the direct correlation
function in the ordered phase using the methods of
\cref{PYordered}. The present authors
also developed a density-functional description
\cite{Paper,Domains} using a relatively crude approximation for the
pair correlation function,
whose results will be reviewed in this paper. In contrast to
\cref{Patey3} and to the simulations this approach enables one to map out whole
phase diagrams which turn out to exhibit an interesting topology
varying with the dipole strength. Other authors have focused on the 
isotropic-ferromagnetic spinodal which corresponds to a divergence of the
dielectric constant, whose position can be estimated using correlation
functions from different integral equation theories
\cite{Hoye,Klapp2}. Finally, we mention perturbation
theory as an alternative approach to determine successfully
liquid-vapor coexistence of Stockmayer fluids with rigid or
polarizable dipoles \cite{Leeuwen,Szalai1,Szalai2,Kriebel}. 

In all theoretical treatments particular care must be taken of the long range of
the dipolar interaction \cite{Paper,OsipovLR} which may give rise to
diverging integrals. Several approaches to handle these problems have
been suggested \cite{Patey3,Paper,OsipovLR}, the most consistent one
being an explicit performance of the thermodynamic limit starting from
a large but finite sample of a given shape \cite{Paper}. For a
homogeneously 
magnetized fluid one finds an energy contribution that depends on the
sample shape. However, this shape dependence disappears if
inhomogeneously magnetized configurations are taken into account
\cite{Griffiths,Domains}. 

These complications do not arise in Heisenberg fluids governed by
short-ranged exchange interactions for which similar
phase transitions and phase diagrams have been found
\cite{TavaresHeisenberg,NijmeijerRev,Oukouiss}. For this model as well as
for the corresponding Ising fluid \cite{NijmeijerIsing} there is
an ongoing discussion concerning the values of the critical exponents
describing the isotropic-ferromagnetic transition
\cite{NijmeijerHSPRL,NijmeijerHeisenberg,NijmeijerIsing}. A Potts
fluid with $q\geq 3$ states exhibits only first-order
isotropic-ferromagnetic transitions \cite{Potts}. For a review
of the properties of dipolar and spin fluids as obtained from
simulations see \cref{NijmeijerRev}.

Experimentally a ferromagnetic liquid phase in dipolar fluids has not
yet been found. The dipole moments of molecular fluids seem to be
too low, though only by a small factor, compared to those that led to
ferromagnetism
 in the simulations. In ferrofluids \cite{Rosensweig},
i.e., colloidal suspensions of small single-domain ferromagnetic
particles, chemical tailoring facilitates higher reduced dipole moments; the relevant dimensionless quantity is $k_B T m^2/\sigma^3$
where $m$ is the dipole moment and $\sigma$ the particle
diameter. However, in the simulations the phase transition always
occurs at high packing fractions that are presently  not attainable in
ferrofluids. Due to the combination of fluidity and strong
paramagnetism ferrofluids already have a wide range of technical applications
that would certainly be enlarged if a ferromagnetic phase could be
established. Electro-rheological and magneto-rheological fluids \cite{ERConf} also
have high technological relevance. In these systems dipole moments are
induced in the colloidal particles by external electric or magnetic
fields and large differences between the permitivities of the solute
and the solvent which then give rise to strong dipolar particle interactions having a
pronounced effect on their rheological properties. However,
spontaneous orientational order in the absence of external fields is
not possible in these fluids.

Ferromagnetic metals  become paramagnetic already far below their
melting temperature. Therefore it came as a surprise when recently
ferromagnetism was detected for the first time in a liquid metal. In
Co$_{80}$Pd$_{20}$ the melting temperature is sufficiently close to
the Curie temperature so that ferromagnetism in the liquid state can
be reached by undercooling \cite{Platzek,Reske,Albrecht}. For the
formation of the ferromagnetic order in these materials short-ranged
exchange interactions play an important role. However, in addition
dipolar interactions are also present and, as in solid ferromagnets,
are essential in forming the domain structure. Therefore for these
systems we expect strong similarities to the behavior of dipolar
fluids as discussed here.

Two other closely related peculiarities of simple dipolar fluids have
attracted much attention recently. Strongly dipolar hard or soft
spheres form chains of head-to-tail aligned particles at very low
densities \cite{WeisCh,LevesqueCh}. The chain formation hinders
the condensation of liquid droplets. Thus a certain amount of
spherically symmetric dispersion interaction is necessary in order to
obtain liquid-gas coexistence in a dipolar fluid \cite{Smit} (but see
also \cref{Tavares}). Alternatively, liquid-gas coexistence of hard
dipolar particles is promoted if they are slightly elongated, which renders the
energies of the head-to-tail and the antiparallel side-to-side
configuration more equal. It has been shown that for a narrow regime
of aspect ratios a liquid phase condensates \cite{McGrother}. The relation
between chaining and condensation has been studied in a number of
theoretical papers \cite{Sear,RoijLett,OsipovCh,Tavares}.

The solid phases of dipolar fluids are less well studied. In
simulations a body-centered tetragonal structure, which has also been
found experimentally in electrorheological fluids \cite{Chen}, as well
as fcc structures with different types of orientational order were
examined \cite{Weis2}, but it was not possible to sort out the
thermodynamically stable phase. Two early density-functional
approaches \cite{Oxtoby,Smithline} did not find a ferromagnetic solid,
but both suffer from algebraic errors \cite{OxtobyErr} and an
incorrect treatment of the long-ranged dipolar interaction
\cite{Solid}. Our own density-functional results for the freezing of
the Stockmayer fluid are presented in Sec.~\ref{solid}.
Ferromagnetism in amorphous dipolar solids is discussed in
Refs.~\cite{Vugmeister,Zhang,AytonAmorph}.

The ferromagnetic liquid should be more stable in fluids of oblate
instead of spherical particles. Indeed a ferromagnetic phase is formed
in a fluid of cut spheres while at lower temperatures a columnar phase
with ferromagnetic order only within the columns is observed
\cite{Ayton}. The relations between nematic and ferromagnetic order in
fluids of
dipolar hard ellipsoids are discussed in Sec.~\ref{diplc} within the
framework of a simple density-functional theory.

Once the bulk phases and their coexistence are established one can
study the interfacial properties which arise if different phases are
brought into spatial contact with each other or with external walls.
These cases can, in
principle, be analyzed using the same density functionals as for the
bulk. This is demonstrated in Sec.~\ref{lginterf} for the 
interface between the isotropic gas and the isotropic liquid which exhibits local orientational order. For dipolar hard
spheres and Stockmayer fluids other theoretical
approaches to this problem can be found in
Refs.~\cite{Eggebrecht,Kasch,TelodaGama}. There is also an extensive
literature concerning more realistic models of specific substances
such as water or methanol using lattice-gas theory
\cite{Matsumoto,Besseling}, integral equation theory \cite{Booth},
density-functional theory \cite{Yang94}, and Monte Carlo simulation
\cite{Sokhan}. 

Non-planar interfaces arise in liquid drops and clusters. Metastable
spherical liquid drops have been studied with density-functional
theory by Talanquer and Oxtoby \cite{Talanquer1,Talanquer2} in
connection with the analysis of liquid nucleation in supersaturated
dipolar gases. In computer simulations of small dipolar clusters \cite{Singer1,Singer2}
interestingly at low temperatures one finds vortexlike orientational order 
similar to the bulk structure of the ferromagnetic phase as discussed
in Sec.~\ref{domains}.

\section{Model and theoretical approach}

We  study the Stockmayer model of a dipolar fluid. It is defined
by the interaction potential 
\begin{equation}
w(\rv_{12},\omega,\omega')=w_{LJ}(\rd)+w_{dip}(\rv_{12},\omega,\omega')
\end{equation}
where $\rv_{12}=\rv-\rv'$ is the interparticle vector and the orientation of
the dipole moments is denoted by $\omega=(\theta,\phi)$. The
Lennard-Jones potential $w_{LJ}(r)=4\epsilon [(\sigma/r)^{12}-(\sigma/r)^6]$ with parameters $\epsilon$ and
$\sigma$ describes the spherically symmetric dispersion interaction as
well as the short-ranged repulsion. The dipolar potential due to point
dipoles embedded in the center of the particles is given by
\begin{equation}
w_{dip}(\rv,\omega,\omega')=-\frac{m^2}{r^3} \left( 3 (\widehat{\bf
m}(\omega) \hat \rv) (\widehat{\bf
m}(\omega') \hat \rv)- \widehat{\bf m}(\omega)\widehat{\bf
m}(\omega') \right)
\end{equation}
where carets denote unit vectors and $m$ is the absolute value of the
dipole moment. A natural extension of this model are mixtures of two
polar or polarizable species which also are important in a number of
technicological applications \cite{HendersonMix,KriebelMix}.

For a start the Stockmayer fluid is treated within the framework of a relatively simple
density-functional theory, which can be derived from an exact
expression for the density functional using the low-density limit
$\exp(-\beta w)$ for the pair distribution function
\cite{Frodl1}. (For more sophisticated choices see \cref{Klapp1}.) The
resulting functional of the one-particle number density
$\hat \rho(\rv,\omega)$ of particles at point $\rv$ and with
orientation $\omega$ is
\begin{eqnarray} \label{DF}
\Omega[\{\hat\rho(\rv,\omega)\},T,\mu]&=&\frac{1}{\beta} \int d^3r
d\omega\, \hat \rho(\rv,\omega)\left[\ln \left(4\pi \Lambda^3
\hat\rho(\rv,\omega)\right)-1\right] +\int d^3r\, f_{CS}(\rho(\rv)) \nn \\
& &{}-\frac{1}{2\beta} \int d^3r d^3r' d\omega d\omega'\,
\hat\rho(\rv,\omega) \hat\rho(\rv',\omega') \Theta(\rd-\sigma)
f(\rv_{12},\omega,\omega') \\
& &{}-\mu \int d^3r d\omega \hat\rho(\rv,\omega). \nn
\end{eqnarray}
Here $\Lambda$ denotes the thermal de Broglie wavelength, $\rho(\rv)=\int d\omega \hat\rho(\rv,\omega)$ is the
orientationally averaged density profile, $f=\exp(-\beta w)-1$ is the
Mayer function, $\beta=1/(k_B T)$ corresponds to the inverse temperature, and
$\mu$ is the chemical potential. The first term in \eqref{DF}
corresponds to the ideal gas free energy while the second term
represents the excess free energy of the hard sphere reference system
in a local density approximation with the Carnahan-Starling free
energy density of hard spheres of packing fraction
$\eta=\frac{\pi}{6}\rho d^3$:
\begin{equation}
  \beta f_{CS}(\rho)=\rho \frac{4\eta-3\eta^3}{(1-\eta)^3}.
\end{equation}
The temperature dependent diameter $d=\int_0^\sigma dr(1-\exp(-\beta
w_{LJ}(r)))$, as given by Barker and Henderson \cite{BHdT}, accounts
for the actually soft repulsion.
The third term in \eqref{DF} describes the contributions to the free
energy due to the long-ranged part of the interactions in a mean-fieldlike manner. The equilibrium density
distribution for a given thermodynamic state minimizes the above
density functional.

For bulk fluid phases the orientationally averaged density $\rho$ is
taken to be
constant and the orientational distribution to be axisymmetric, so
that with
\begin{equation}
\hat\rho(\rv,\omega)=\rho(\rv) \alpha(\rv,\omega), \qquad
\int d\omega\,\alpha(\rv,\omega)=1,
\end{equation}
in general, here one has
\begin{equation} \label{densansatz}
\hat\rho(\rv,\omega)=\rho
\alpha(\cos\theta)=\frac{\rho}{2\pi}\sum_{l=0}^\infty \alpha_l
P_l(\cos\theta).
\end{equation}
In the isotropic phase $\alpha(\cos\theta)=1/(4\pi)$,
i.e., $\alpha_l=\delta_{l0}/2$, while in the ferromagnetic phase also the
higher expansion coefficients $\alpha_{l\geq 1}$ take on non-zero
values. If one inserts the expansion in \eqref{densansatz} into the density
functional and employs a suitable expansion of the Mayer function into
rotational invariants of the three spatial angles $\omega$, $\omega'$
and $\omega_{12}$, one encounters a conditionally convergent integral
that is proportional to $\alpha_1^2$ \cite{Paper}. This is a consequence of the
slow decay of the dipolar potential (and therefore of the
Mayer function). The value of this integral for fluid volume
$V\to\infty$ depends on the shape of the sample adopted for this
limiting process. This integral was determined for ellipsoidal shapes of
arbitrary aspect ratio $k$ and shown to be related to the
demagnetization factor known from classical magnetostatics. On the other hand
it has been shown by general arguments that the free energies and the
phase diagrams of dipolar systems in zero field do not depend on the sample shape 
\cite{Griffiths}. This apparent contradiction is reconciled by the
observation that the true equilibrium state of the ferromagnetic
liquid corresponds to such a shape-dependent inhomogeneous
configuration that in the resulting free energy the shape dependence
drops out. Using a general spatially inhomogeneous
orientational distribution, within the present theory we have confirmed
that the equilibrium free energy indeed does not depend on the sample
shape and is the same as that of the homogeneous state in 
a needle-shaped volume ($k\to\infty$) \cite{Domains}. Therefore the
phase diagram determined for this special case is valid for all shapes
 keeping in mind that for a general shape the ferromagnetic phase exhibits a
nontrivial spatial structure. Details of this structure are discussed
in the next section. We remark that these inhomogeneities are surpressed by
the so-called ``tin-foil'' boundary conditions that are usually applied in
computer simulations, i.e., by surrounding the sample with an
infinitely permeable material (a perfect conductor in the electric
case, respectively). The free ``magnetic charges'' in the surrounding material
cancel the demagnetization field which leads to shape independence
also for homogeneous magnetization. In the following we only consider
the more realistic open boundary conditions.

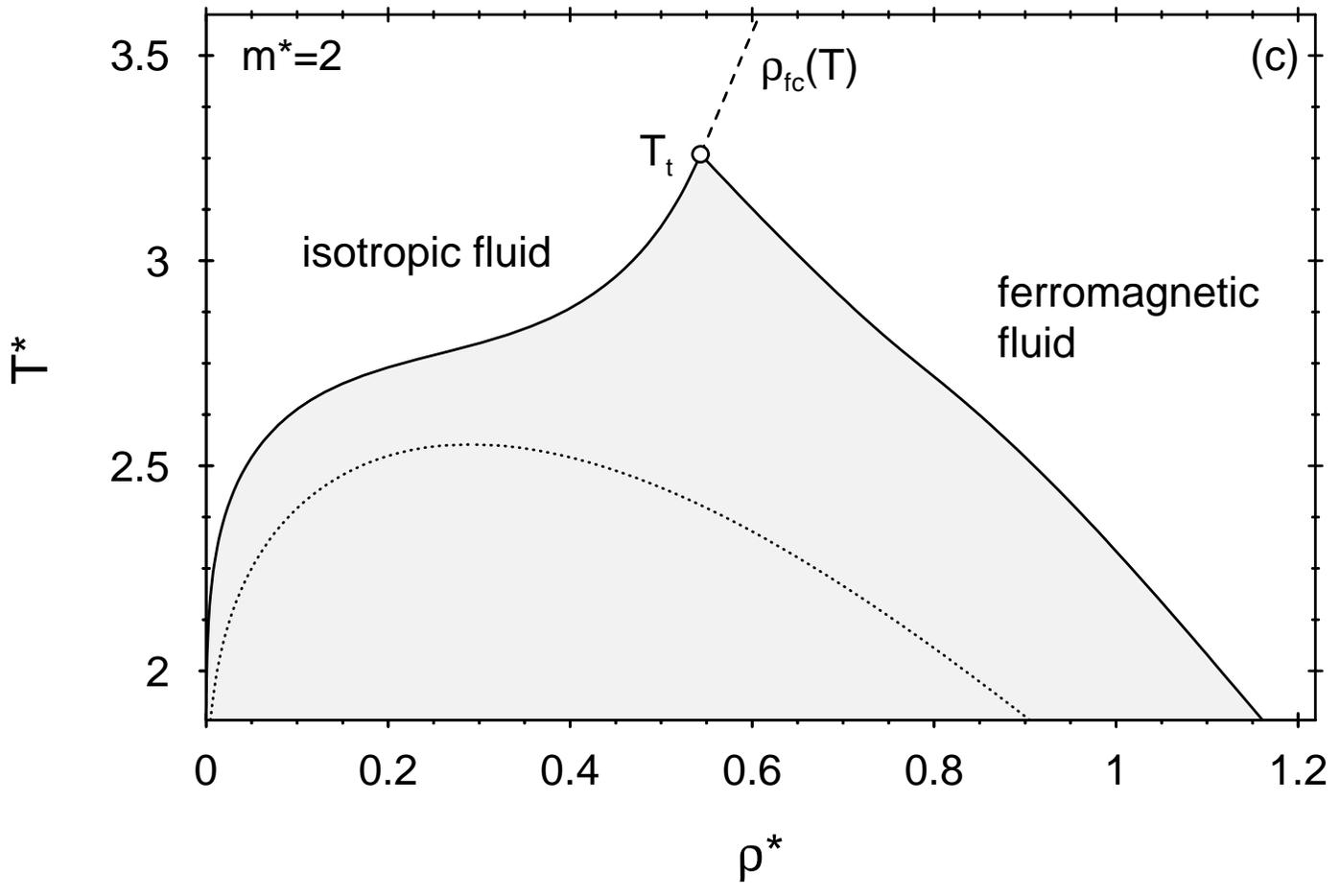
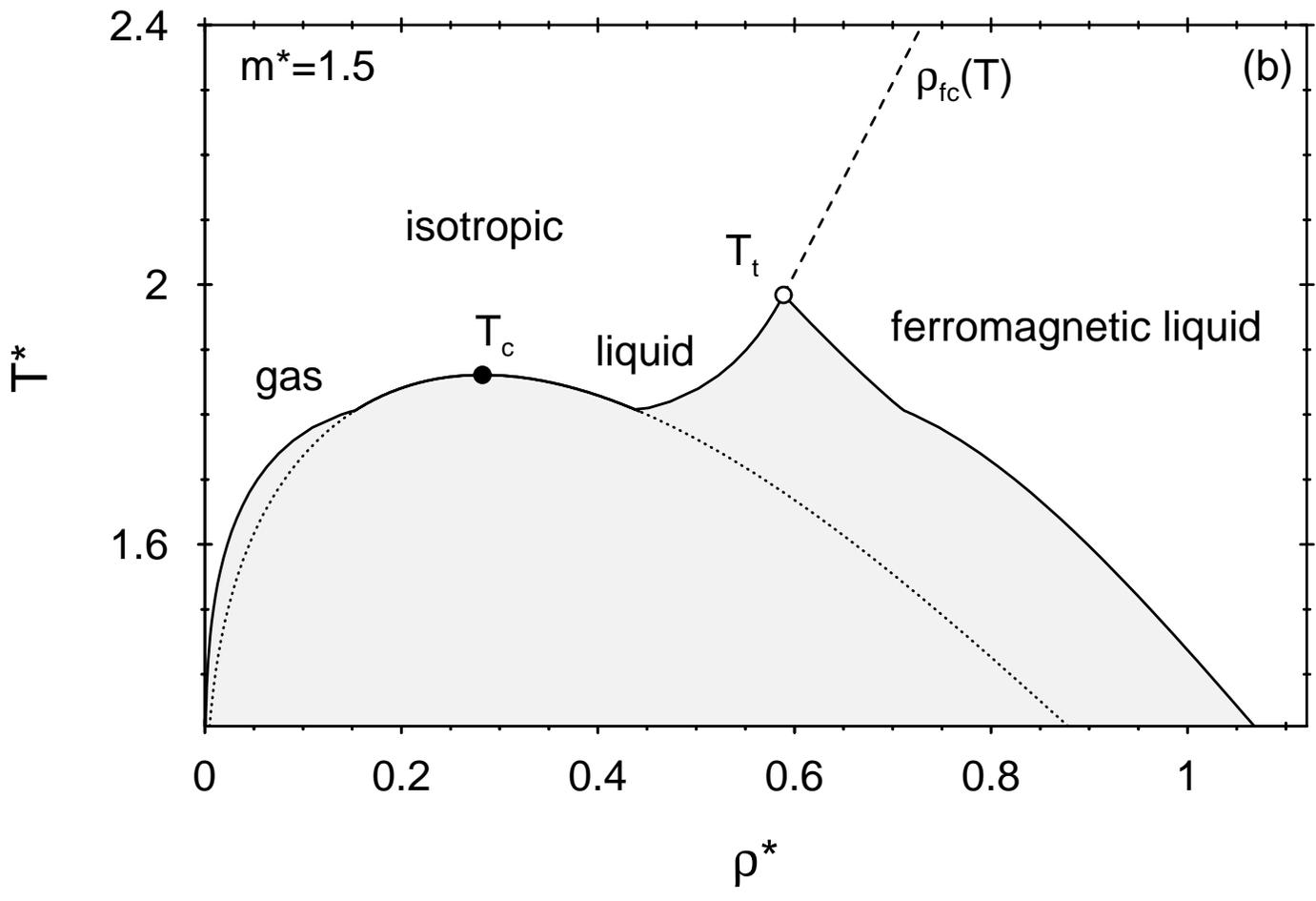
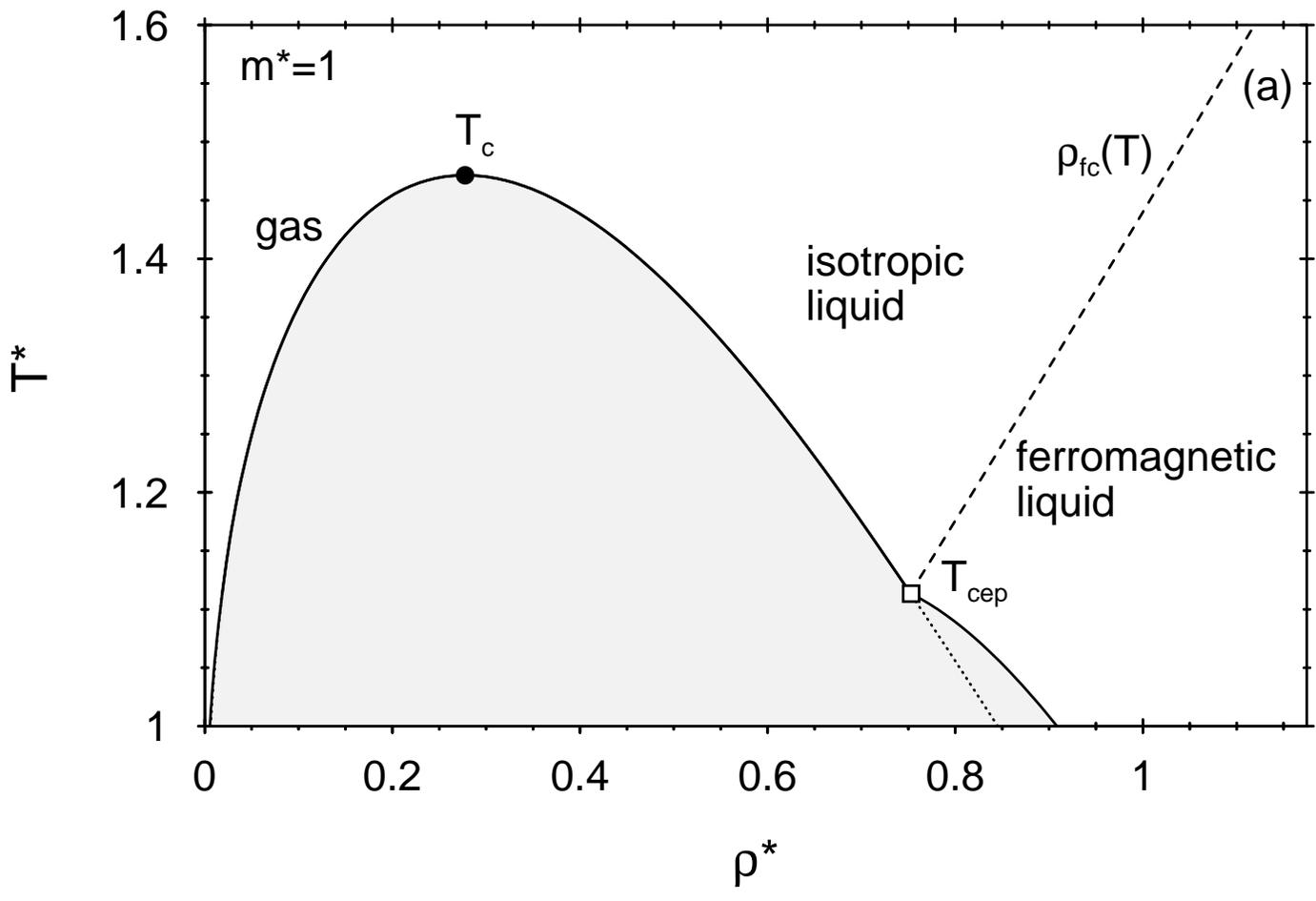
\begin{figure}[htbp]
\caption{Phase diagrams of the Stockmayer fluid for three different
values of the reduced dipole moment $\ms$, including the ferromagnetic
liquid phase. The dashed line $\rho_{fc}(T)$ denotes a line of
critical points, $T_c$ is the gas-liquid critical point, and the
shaded areas represent two-phase regions. In part (a) the critical
line ends at the coexistence line in a critical end point $T_{cep}$,
while in  (b) and (c) a tricritical point $T_t$ arises, below which the
isotropic-ferromagnetic transition becomes first order. The dotted
lines represent the metastable parts of the coexistence curve of the
isotropic phases.}
\label{fig:pdfluid}
\end{figure}

The properties of the Stockmayer fluid can be described in terms of the
following dimensionless quantities:
\begin{equation}
 \ts=k_B T/\epsilon,\qquad \rhos=\rho\sigma^3,\qquad
 \ms=m/\sqrt{\sigma^3\epsilon}.
\end{equation}
Figure~\ref{fig:pdfluid} displays the phase diagrams in the
temperature-density plane as obtained from the present theory for
three different values of the reduced dipole moment $\ms$. At low
dipole moments there is a large coexistence region of the isotropic
gas and liquid phases and a second-order transition to the
ferromagnetic liquid along a line of critical points. This line ends
at the coexistence curve in a critical end point. Upon increasing the
dipole moment the isotropic-ferromagnetic transition shifts to lower
densities and higher temperatures and a tricritical point emerges
below which the transition becomes first order. Moreover all three
fluid phases coexist at a triple point. At even higher dipole moments
the gas-liquid coexistence becomes metastable and is preempted by the
isotropic gas - ferromagnetic liquid transition, so that one is left with only two fluid
phases. A similar series of phase diagrams has been obtained from a
phenomenological theory \cite{WZ}, while in simulations to date
neither the order of the transition nor the topology of the phase
diagram have been mapped out.

\section{Inhomogeneous magnetization} \label{domains}

The arguments leading to the proof of the shape independence mentioned
above provide also a characterization of the actual
inhomogeneous magnetization configuration for a given sample shape.
Locally it should exhibit the same orientational distribution that has
been determined for the needle shaped sample, but with a spatially
varying preferential direction. This variation should correspond to a
magnetization field $\Mvec(\rv)$ which on the scale of the system size
has zero divergence and a vanishing normal component at the surface
\cite{Domains}. However, for a given shape these conditions do not determine
$\Mvec(\rv)$ uniquely and we are not aware of a systematic
method to construct fields consistent with them. Therefore we pursued
an alternative approach by computing this field via a
quasi-free minimization of the density functional. A cubic
discretization lattice with lattice constant $a$ and lattice vectors
$\Rv$ is superimposed on a cubic sample which we have chosen as an example. From
\eqref{DF} one derives the following approximation for the free energy
difference $\Delta F$ between the isotropic and ferromagnetic phases
\cite{Simann}: 
\begin{equation} \label{DFsimann}
 \beta \Delta F=\rho a^3 \sum_{\Rv} \sum_{n=1}^\infty
 \frac{(\sqrt{12\pi} M(\Rv))^{2n}}{(2n-1)2n(2n+1)}
 +\half\rho^2 a^6 \sum_{\Rv,\Rv'} \sum_{i,j} M_i(\Rv) w_{ij}(\Rv-\Rv')
 M_j(\Rv')
\end{equation}
with the interaction tensor
\begin{equation}
 w_{ij}(\Rv)=\Theta(R-\sigma)\left[f_{110}(R)
 \delta_{ij}+f_{112}(R)(\delta_{ij}-3 \hat R_i \hat R_j)\right].
\end{equation}
The functions $f_{110}$ and $f_{112}$ are certain expansion
coefficients of the Mayer function that are known analytically. To
lowest order in $m^2$ one has $f_{110}\sim m^6$ and $f_{112}(r)\sim
(1-e^{-\beta w_{LJ}}) m^2/r^3$ so that $w_{ij}(\Rv-\Rv')$ reproduces
the dipolar interaction between the points $\Rv$ and $\Rv'$. This
expression is minimized with respect to the values of $\Mvec(\Rv)$ at
all lattice points using the simulated annealing algorithm
\cite{NumericalRecipes}. Up to $N=24$ points for each spatial dimension are used which
amounts to $3\times 24^3=41472$ variables.

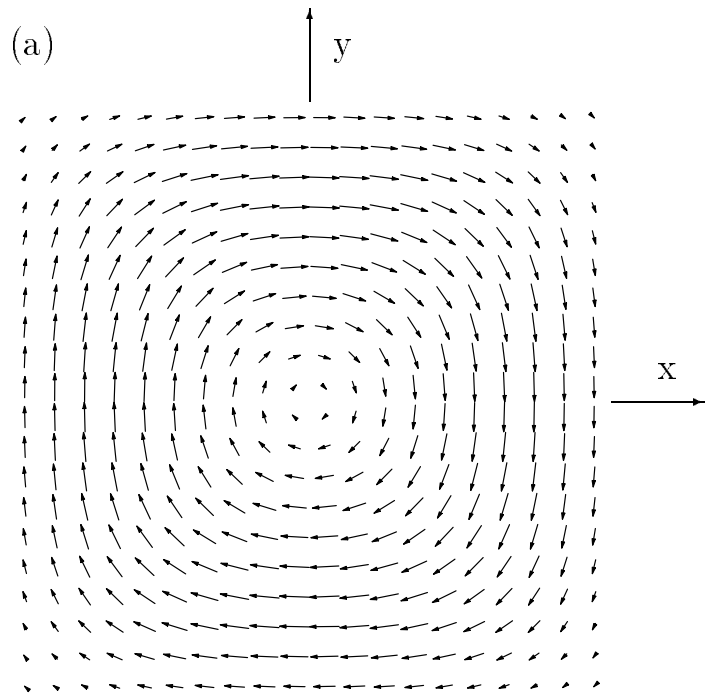
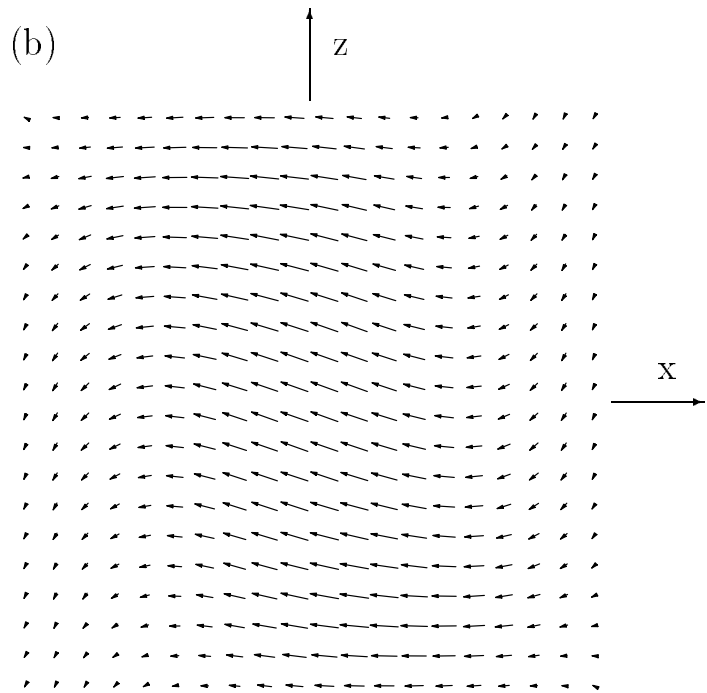
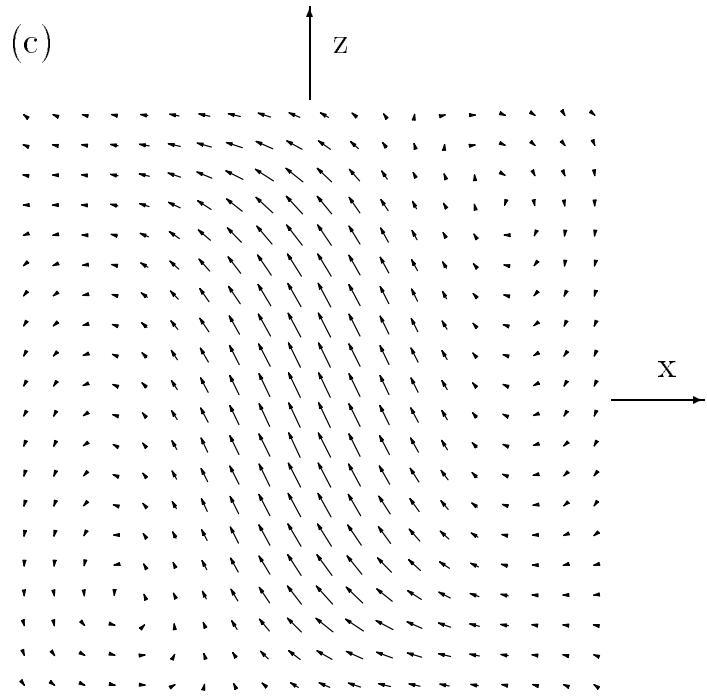
\begin{figure}[htbp]
\caption{Magnetization structure of a ferromagnetic liquid in a
    cubic volume with edge length $L/\sigma=7.2$, $20^3$ mesh points,
    and $T^\ast=2.25$. The three parts of the
    figure represent sections (a) for fixed $z=0.025 L$, (b) fixed
    $y=-0.075 L$, and (c) fixed $y=-0.225 L$; the origin is at the
    center of the cube. The arrows represent projections of the local
    magnetization at their midpoint onto the section plane. The $z$
    axis is the vortex axis. Part (c) demonstrates that in the
    vicinity of this axis the magnetization ``escapes'' from the
    vortex plane.}
\label{fig:structure}
\end{figure}

Results for the thermodynamic parameters $\ms=1.5$, $\ts=2.25$, and
$\rhos=0.94$ are shown in \abbref{fig:structure}. One of the three
perpendicular cubic axes is chosen spontaneously as a vortex axis
around which the magnetization field circulates. As one can infer
from the sections parallel to the vortex axis (i.e., the $z$ axis) in
Figs.~\ref{fig:structure}(b) and (c) $\Mvec(\rv)$ lies essentially in
the vortex plane except close to the axis where it ``escapes'' into
the third dimension in order to avoid a line singularity. Since the
magnetization is approximately parallel at the surface two point
defects at the centers of the top and bottom face are
inevitable. The absolute value $M(\rv)$, which in
contrast to the common assumption in micromagnetic calculations
\cite{Micromagnetics} is not fixed at a constant value, is strongly
reduced within the cores of the point defects whereas it is
approximately constant throughout the rest of the volume with the
exception of thin surface layers. At higher temperatures in the
vicinity of the phase transition to the isotropic liquid a reduction
of $M(\rv)$  occurs also along the whole vortex axis.

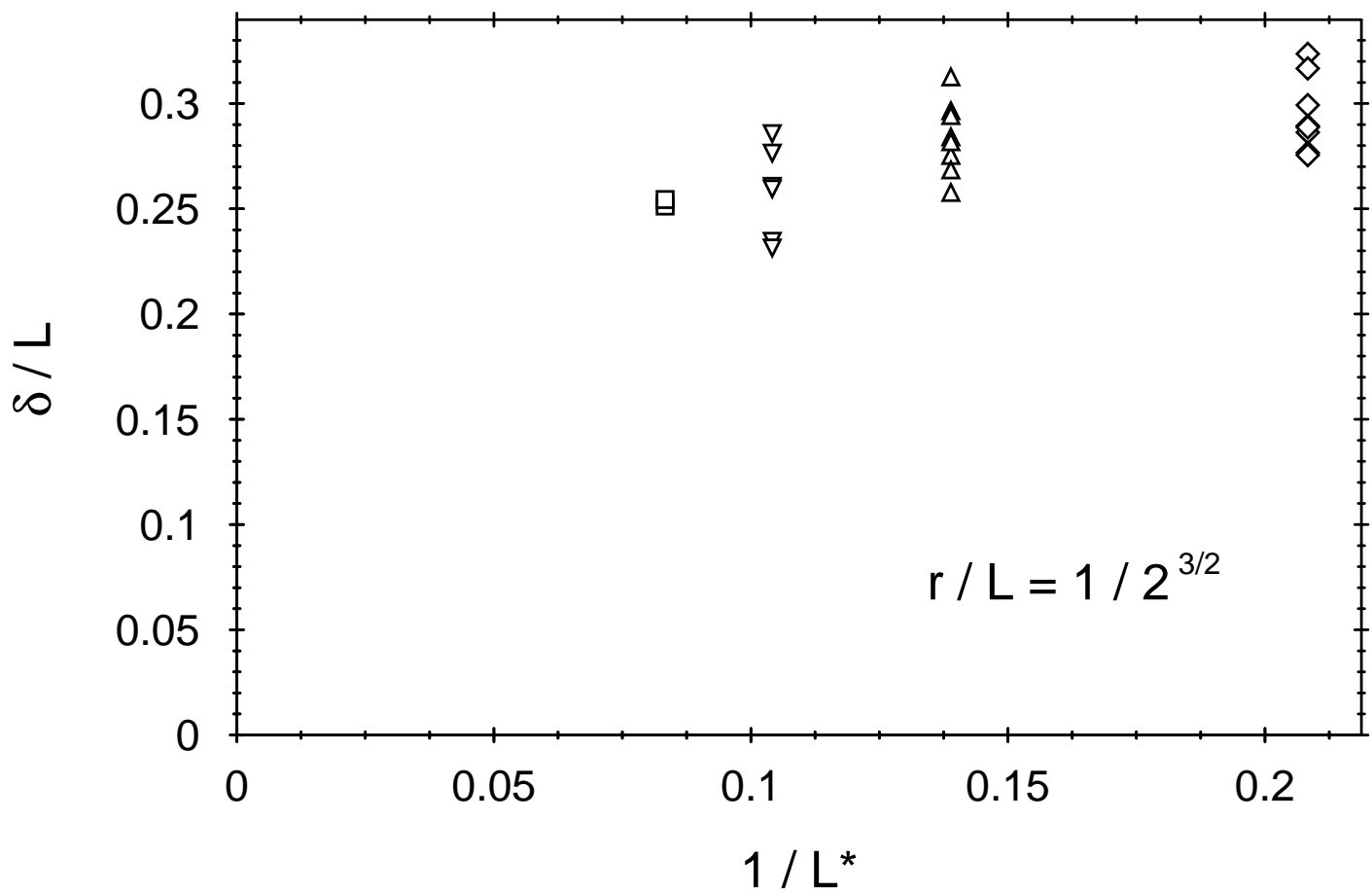
\begin{figure}[htbp]
\caption{The domain wall thickness $\delta$ in a cubic sample measured
within the plane $z=0$ at distance $r=L/2^{3/2}$ from the center,
i.e., at half distance between the center and the edges. The different results
for the same value $L^\ast=L/\sigma$ correspond to different numbers of mesh
points used for the minimization. The extrapolation to $1/L=0$
suggests a finite value of $\delta/L$ in the thermodynamic limit.}
\label{fig:thickness}
\end{figure}

The structure in \abbref{fig:structure} can also be considered to
consist of four triangular domains with four broad 90$^o$ domain
walls along the diagonals. These walls are clearly induced by the cube
edges and would probably be absent in a cylindrical sample. We have
computed a suitably
defined wall thickness $\delta$ \cite{Domains}  for
different sample sizes $L$. In \abbref{fig:thickness} we show the
results as function of $1/L$ up to the largest systems with
$L/\sigma=12$. The various points for the same $L$ correspond to
different values of $N$ and thus give an indication of the error
induced by the discretization. The data are consistent with the
conjecture $\delta/L\to{\rm const}$ for $L\to\infty$ which would mean
that the wall thickness is determined by the system size \cite{DeGennes1}. This
contrasts with the finite wall thickness in solid ferromagnets that is
caused by the existence of easy axes induced by the lattice
anisotropy. The whole magnetization configuration fulfills, to a good
approximation, the scaling relation
\begin{equation}
 \Mvec(r/\sigma,L/\sigma)\approx\Mvec^{(0)}(\rv/L)
\end{equation}
when different sizes within the accessible regime are compared. We
conjecture that this holds  also in the thermodynamic limit.

The average magnetization of the sample $\av{\Mvec}=V^{-1} \int d^3r
\Mvec(\rv)$ is approximately zero although $|\Mvec(\rv)|\neq 0$. The positive
contribution to $\av{M_z}$ from the ``escape'' near the vortex axis is
compensated by negative contributions from the regions close to the
edges. However, a finite value of $\av{M_z}$ is expected for cylindrical or
spherical \cite{Aharoni} samples.

\section{External field}

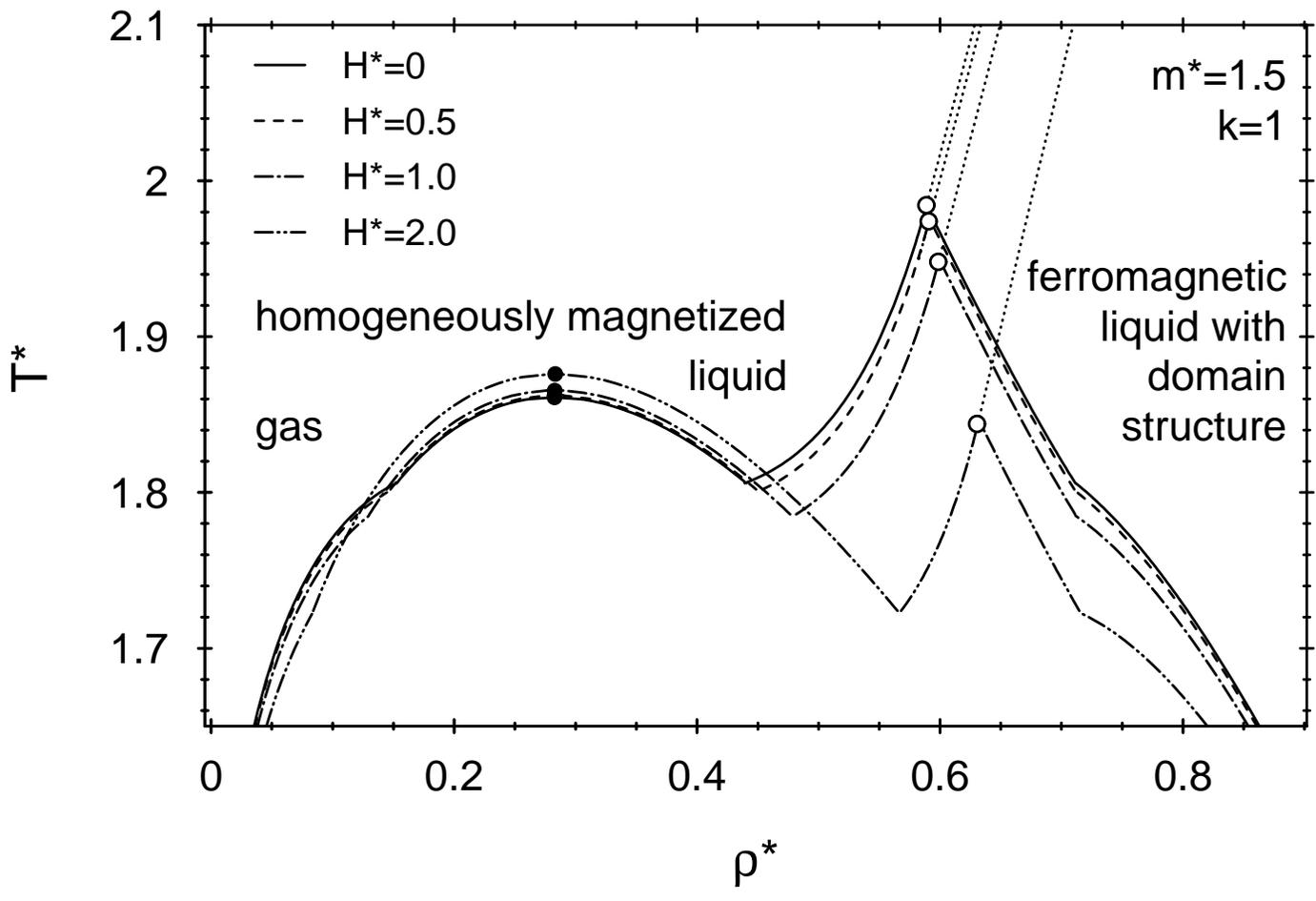
\begin{figure}[htbp]
\caption{Phase diagrams for $\ms=1.5$ and various fixed values of the
reduced external field strength $H^\ast=H\protect\sqrt{\sigma^3/\epsilon}$ for
a sample with aspect ratio $k=1$ (i.e., demagnetization factor
$D=1/3$). For nonzero fields the inhomogeneous domain structure is
partly retained at high densities and low temperatures. The magnetic
field shifts the  transition line of critical points (dotted curve)
and the tricritical point (open circle) towards
higher densities and lower temperatures, whereas the gas-liquid
critical temperature (full circle) is slightly increased.}
\label{fig:phasdh}
\end{figure}

An external magnetic field tends to rotate all dipoles into the field
direction and thus to destroy any inhomogeneity of $\Mvec(\rv)$. Let
us first consider sample shapes which induce a planar zero-field
structure. As we have seen, this is approximately true for the cube,
but most probably also for a torus \cite{Wojtowicz}. We consider the
case that the field is applied
perpendicularly to the magnetization plane. If one assumes that the
angle by which the magnetization is rotated out of plane is the same
throughout the sample one can still treat the inhomogeneous
problem quasi-analytically \cite{Domains}. Upon increasing the external
field the parallel component of $\Mvec$ first increases linearly and
finally saturates while the perpendicular components decrease and
vanish with a square root singularity at a critical field strength
$H_c$ above which the sample is homogeneously magnetized. One can also
determine phase diagrams at fixed external fields which now do depend
on the sample shape via the demagnetization factor $D$ \cite{Domains}. The results
for a cubic or spherical sample ($D=1/3$) with $\ms=1.5$ are shown in
\abbref{fig:phasdh}. The critical line between homogeneous isotropic
and inhomogeneous
ferromagnetic fluids in zero field turns into a critical line between
homogeneously and inhomogeneously magnetized fluids. With increasing
field strength this line together with the tricritical point shifts to
higher densities and lower temperatures. (Only for a needle-shaped
sample ($D=0$) this line disappears and the tricritical point turns
into a second critical point because all phases are homogeneous.) The
small shift of the liquid-gas critical temperature is proportional to
$H^2$ for small fields in accordance with the results of perturbation
theory \cite{Boda}. The influence of an external field on the
liquid-gas coexistence has also been studied in Refs.~\cite{Woodward,Grest,Boda96}.

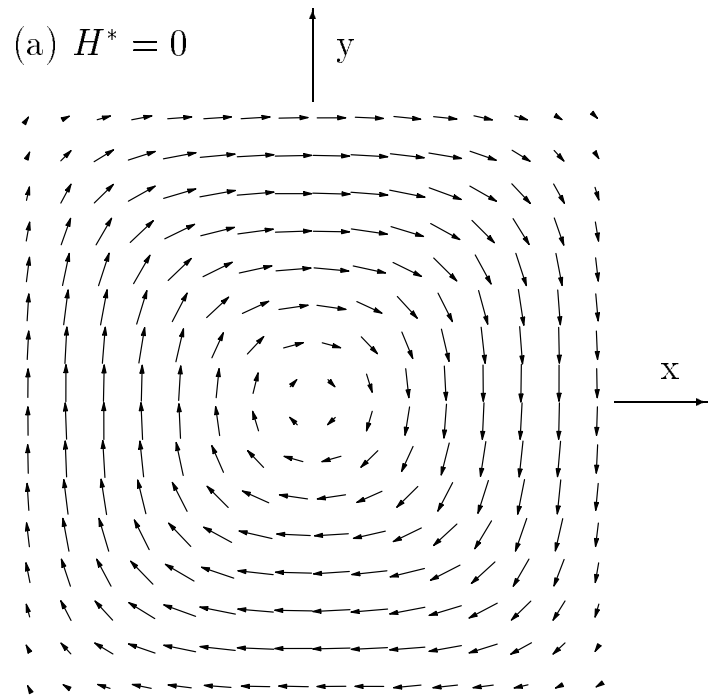
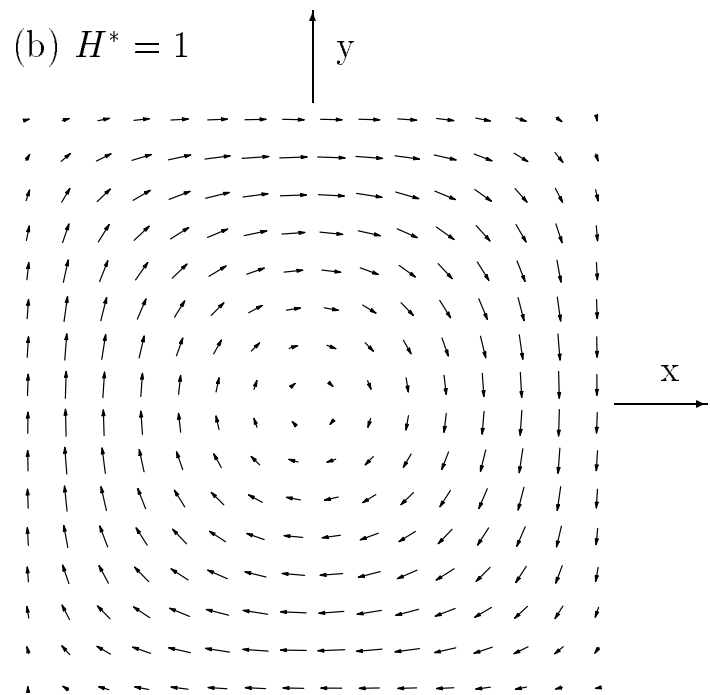
\begin{figure}[htbp]
\caption{Sections of the magnetization structure orthogonal to the field direction   for
$L/\sigma=9.6$ and $N=16$ at $z/\sigma=-0.9$ (a) without and (b) with an
external field $H^\ast=H \protect\sqrt{\sigma^3/\epsilon}=1$ applied
in  the $z$
direction. The scale factor which determines the lengths of the arrows
is the same  in both parts of this figure and also in
\abbpref{fig:structh2} below. The transversal components of the
magnetization are reduced by the
field but the overall feature of the inhomogeneous structure is preserved.}
\label{fig:structh1}
\end{figure}

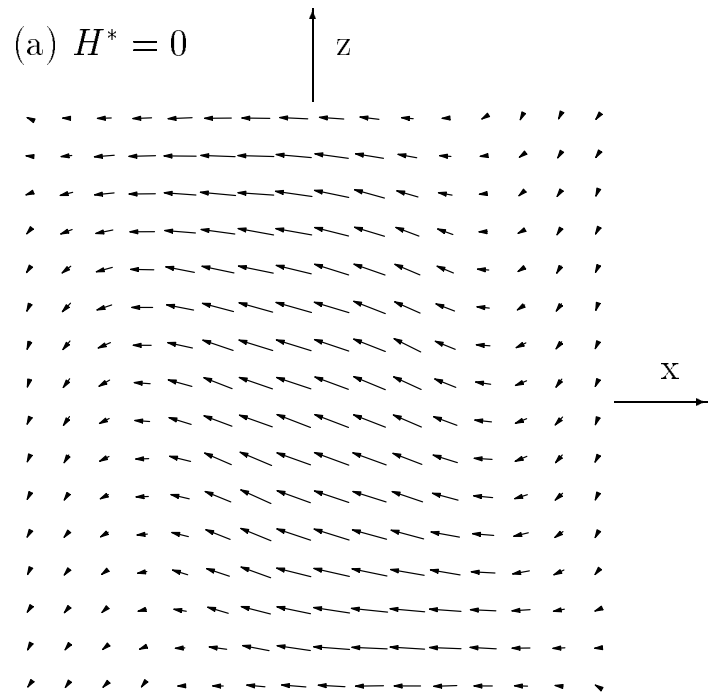
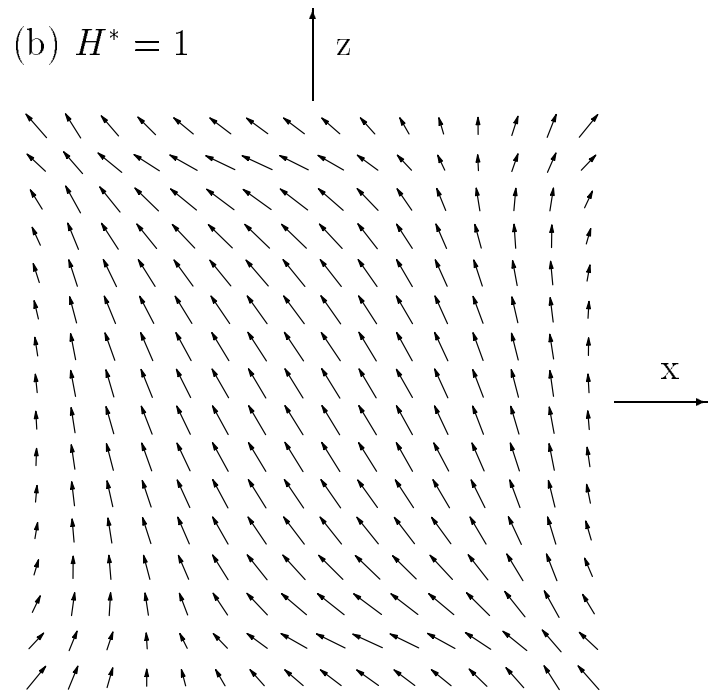
\begin{figure}[htbp]
\caption{Vertical sections of the
same magnetization structures as in \abbpref{fig:structh1} perpendicular to the $y$ axis and thus
parallel to the field direction   for
$y/\sigma=-2.1$. At this distance of the plane from the center the
$z$ component of the magnetization in zero field is only small, while the external field induces a large
$z$ component everywhere.}
\label{fig:structh2}
\end{figure}

If the external field is not perpendicular to the
magnetization plane,  for small fields the structure is
still similar to the zero field result. On the other hand, for
example in a cube there are six degenerate zero-field solutions. By
using the free minimization procedure described in the previous
section one can address the question which of them is selected by a
small external field parallel to one of the cube axes. It turns out
that a slightly perturbed vortex structure with its axis parallel to
the field is favored \cite{Simann}. The further evolution of the
structure with increasing field strength takes place as described
above. Again the azimuthal magnetization component around the field
direction vanishes at a critical field strength $H_c$, leaving an
almost homogeneously magnetized sample for $H>H_c$. A comparison
between the structures in zero field and in a finite field below $H_c$
is given in Figs.~\ref{fig:structh1} and \ref{fig:structh2}. It is
interesting to consider
 the corresponding behavior of $\Mvec(\rv)$ for a {\it
spherical} sample which possesses infinitely many equivalent
directions. Fredkin and Koehler \cite{Fredkin12} have
performed micromagnetic calculations for solid spherical
particles. For low lattice anisotropy they find again a vortex state,
but the vortex axis rotates away from the field direction as $H$ vanishes.

\section{Solid phases} \label{solid}

It is important to obtain an estimate for the position of the freezing
transition in the phase diagram in order to know whether the formation
of the ferromagnetic liquid is preempted by freezing. In order to
get a reasonable description of the solid phase we have chosen a more
sophisticated 
density functional \cite{Solid}. The hard sphere reference fluid is
now treated within the modified weighted density approximation (MWDA)
introduced by Denton and Ashcroft \cite{Denton} which gives very accurate
results for the freezing properties of hard spheres. The perturbative
contribution is changed in a way that reproduces  the results of
Curtin and Ashcroft \cite{CA} in the limit of a Lennard-Jones fluid
but also includes a contribution from the Mayer function of the
dipolar potential as in \eqref{DF}. The number density is parametrized as
\begin{equation} 
 \hat\rho(\rv,\omega)=\rho(\rv)\alpha(\omega)=\frac{1}{2\pi}
 \left(\frac{\lambda}{\pi}\right)^{3/2} \sum_{\Rv} e^{-\lambda
 (\rv-\Rv)^2} \sum_{l=0}^\infty \alpha_l P_l(\cos\theta)
\end{equation}
where $1/\sqrt{\lambda}$ is the width of the density peaks at the
lattice points $\Rv$. The shape dependent contribution to the density
functional has been calculated also for a solid structure. For cubic
lattices it has been shown to be
the same as that of a liquid.

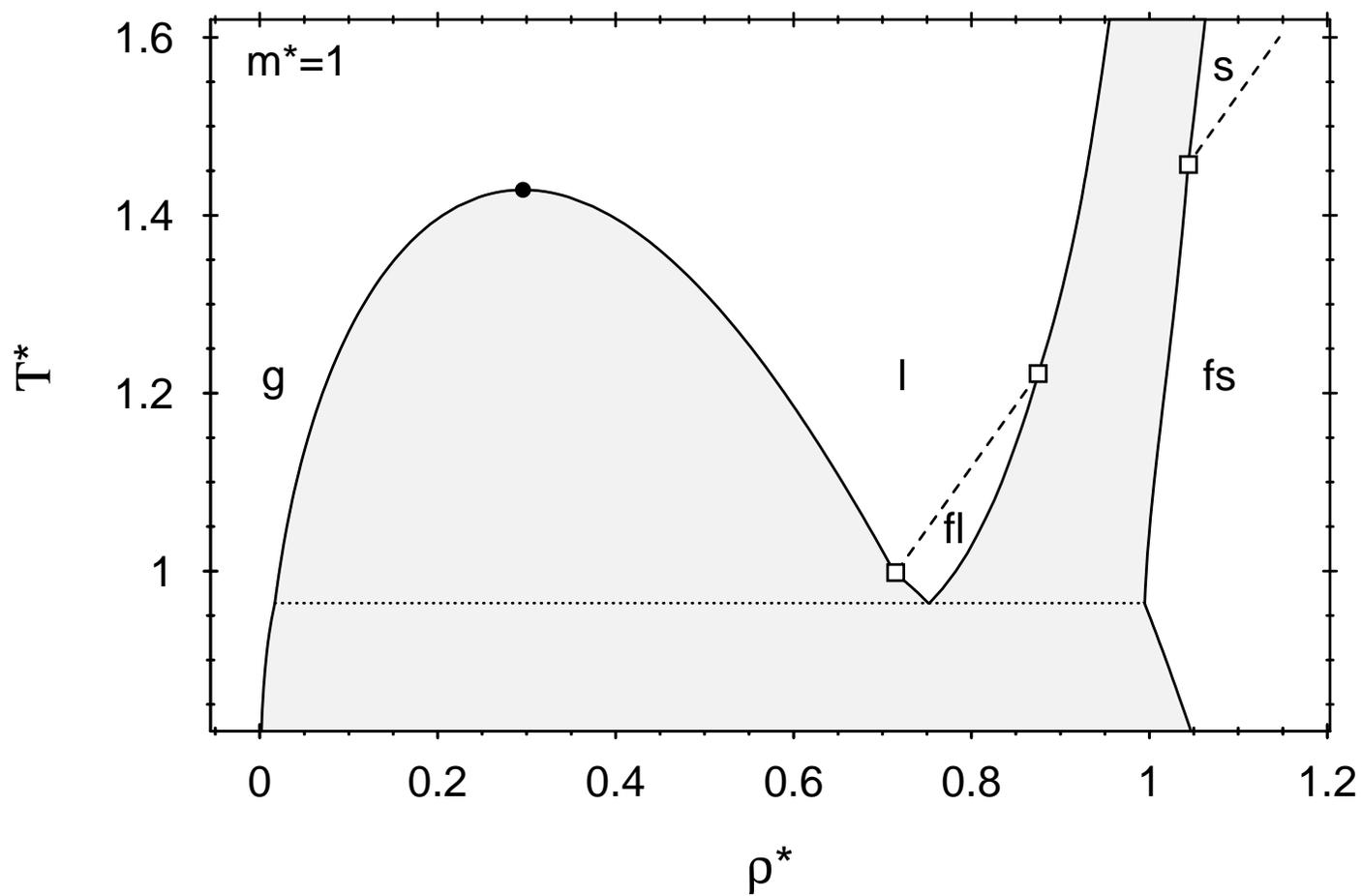
\begin{figure}[htbp]
\caption{Phase diagram for $\ms=1$ including the solid phase as obtained from the generalized
density-functional theory. It
encompasses five phases: the isotropic gas (g) and liquid (l), the
ferromagnetic liquid (fl), the ferromagnetic solid (fs), and the solid
without orientational order (s). Two-phase regions are shaded, the
dashed lines indicate continuous phase transitions, and the
dotted line denotes the triple point. Critical end points are denoted
by squares.}
\label{fig:solidm1}
\end{figure}

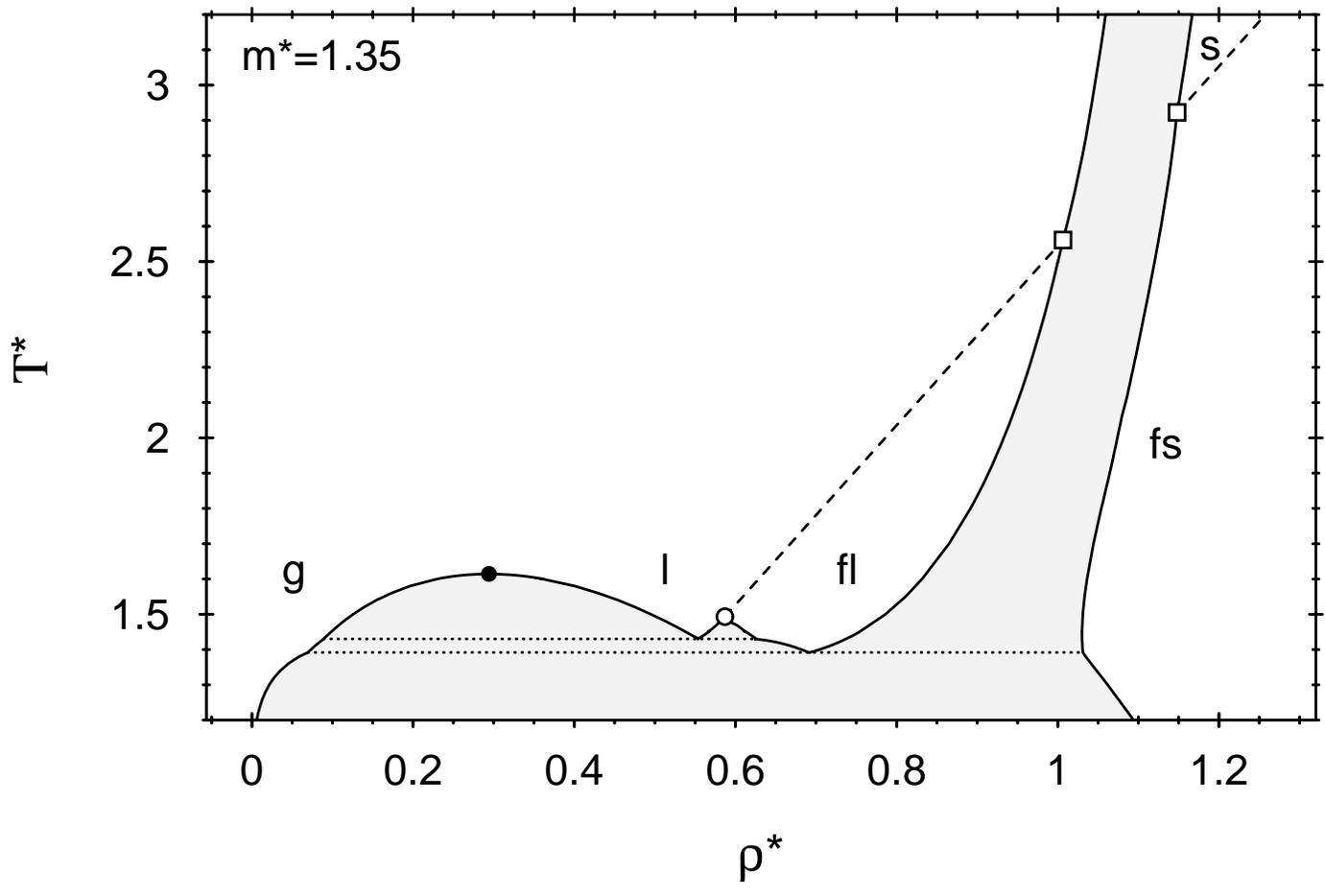
\begin{figure}[htbp]
\caption{The same as \abbpref{fig:solidm1}, but for $\ms=1.35$. In
addition to the features of \abbref{fig:solidm1} a tricritical point
(open circle) and 
two triple points (dotted lines) occur.}
\label{fig:solidm135}
\end{figure}

 For the considered parameter values ($\ms\leq 2$) the fcc lattice was
found to be more stable than the bct (body centered tetragonal)
structure observed in simulations \cite{Weis2}. The phase diagrams
obtained by this approach (Figs.~\ref{fig:solidm1} and
\ref{fig:solidm135}) comprise five different phases, among them a
ferromagnetic solid (fs) magnetized in the (111) direction, and a
plastic solid (s) without orientational order. The transition between
the liquid phases with and without orientational order extends into
the solid region. For dipole moments larger than $\ms\approx 1$ there
is indeed a region with a stable ferromagnetic liquid (fl)
phase. However, based on a different density-functional theory
Klapp and Forstmann \cite{Klapp1} have claimed recently that the latter phase is
metastable with respect to the solid for all dipole moments, at
variance with the simulation results. Thus even the qualitative
features of the Stockmayer fluid phase diagram are not yet finally
settled.

\section{Nonspherical particles} \label{diplc}

In fluids consisting of nonspherical particles an orientationally
ordered liquid phase may already occur in the absence of dipolar
interactions. This nematic phase can be stabilized by short-ranged steric interactions between sufficiently elongated or disklike
particles. It has a higher symmetry than a ferromagnetic liquid as
particle orientations parallel and antiparallel to the preferential
direction have identical probabilities. It is interesting to study the
interplay of these two ordering mechanisms in a system of nonspherical
dipolar particles. Specifically we have considered hard ellipsoids and hard
spherocylinders with a central longitudinal dipole moment
\cite{Diplc}. In this case, the reference  part of the density
functional coincides with the hard core part which
is well approximated by the so-called decoupling approximation that amounts to
replacing the direct correlation function by that of a hard-sphere
fluid scaled by the distance of closest approach for given
orientations of the particles and their joining vector. The dipolar
interaction is again treated perturbatively
within the low-density approximation that yields
an expression  bilinear in the densities with the Mayer function kernel
as in \eqref{DF}. 

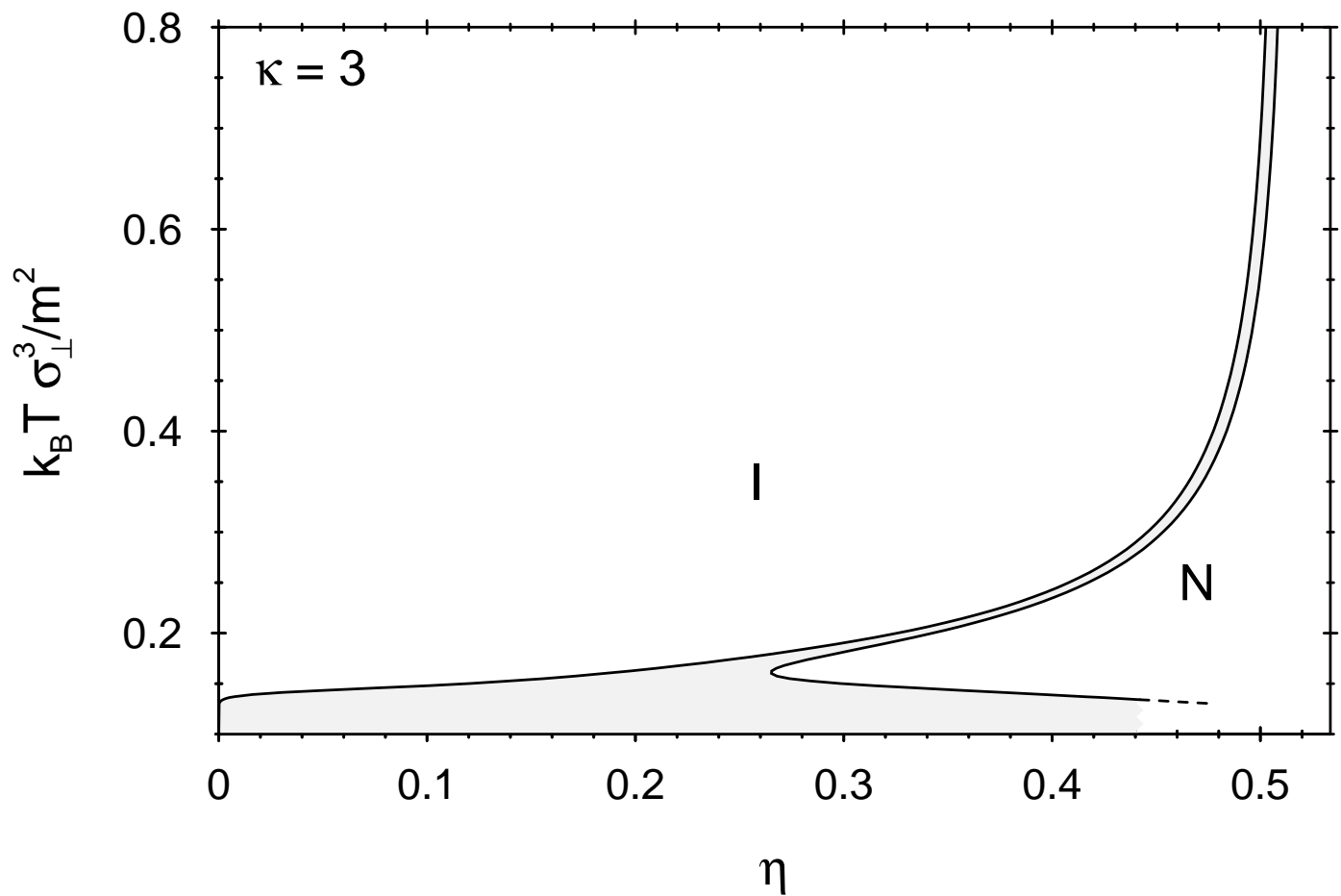
\begin{figure}[htbp]
\caption{Phase diagram of hard ellipsoids of revolution with a
longitudinal dipole moment and aspect ratio $\kappa=3$. $\eta$ is the
packing fraction. At high temperatures or small dipole moments the
transition between the isotropic (I) and the nematic (N) phases is
weakly first order, but it widens into a gas-nematic coexistence at
low temperatures. Possible smectic and solid phases cannot be
described within the present theory.}
\label{fig:pdk3}
\end{figure}

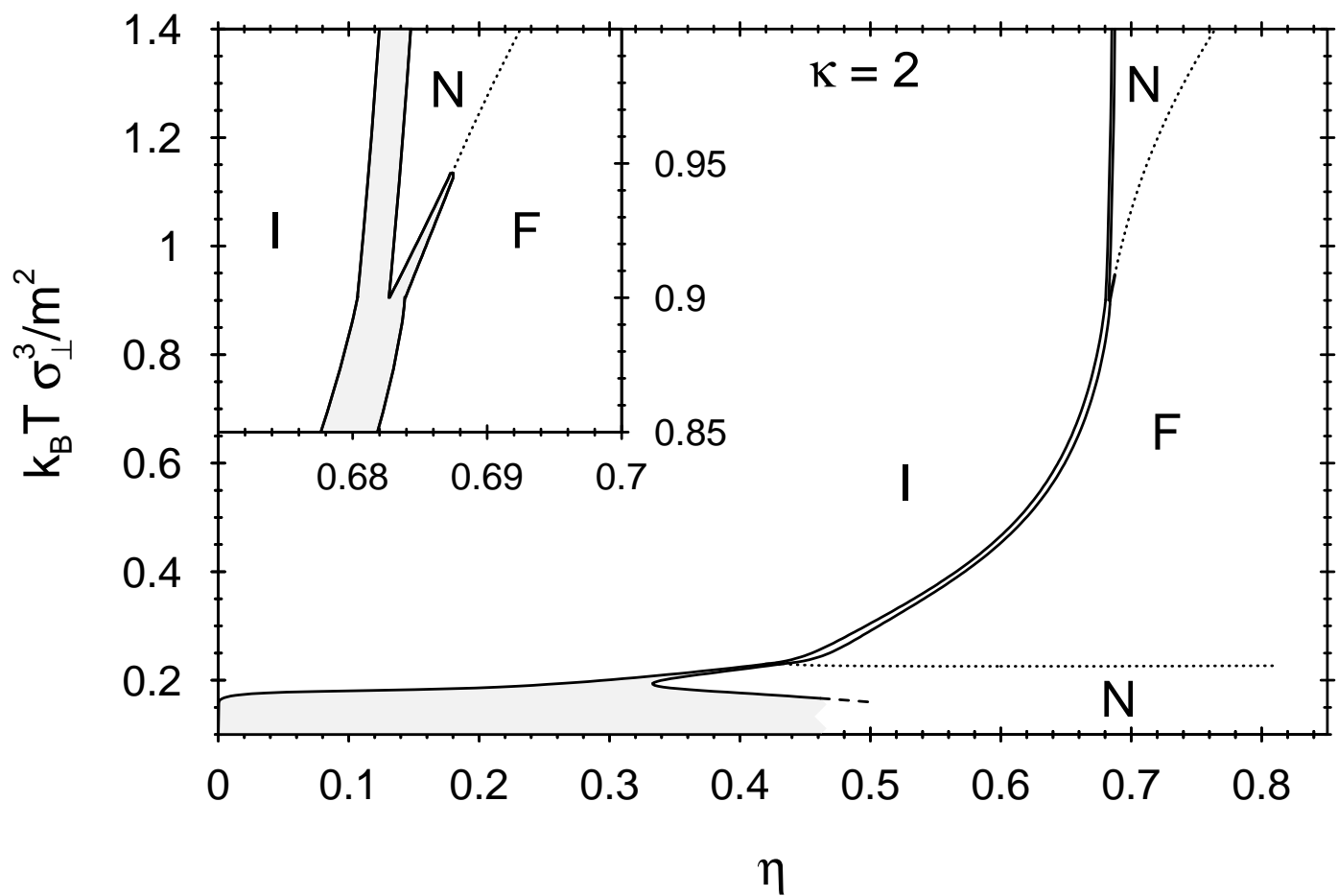
\begin{figure}[htbp]
\caption{The same as \abbpref{fig:pdk3} but for aspect ratio
$\kappa=2$. Besides the isotropic (I) and nematic (N) phases a
ferromagnetically ordered liquid (F) occurs in the intermediate
temperature range. The dotted  and solid lines denote second- and
first-order phase transitions, respectively. The density gap of the
I-N transition at $\eta\simeq 0.68$ cannot be resolved on the present
scale. The inset shows
that in a narrow temperature range the high temperature continuous N-F
transition is turned into a weakly first-order transition generating a
tricritical point and an I-N-F triple point. The reentrant nematic
phase at low temperatures is probably an artefact of the present theory.}
\label{fig:pdk2}
\end{figure}

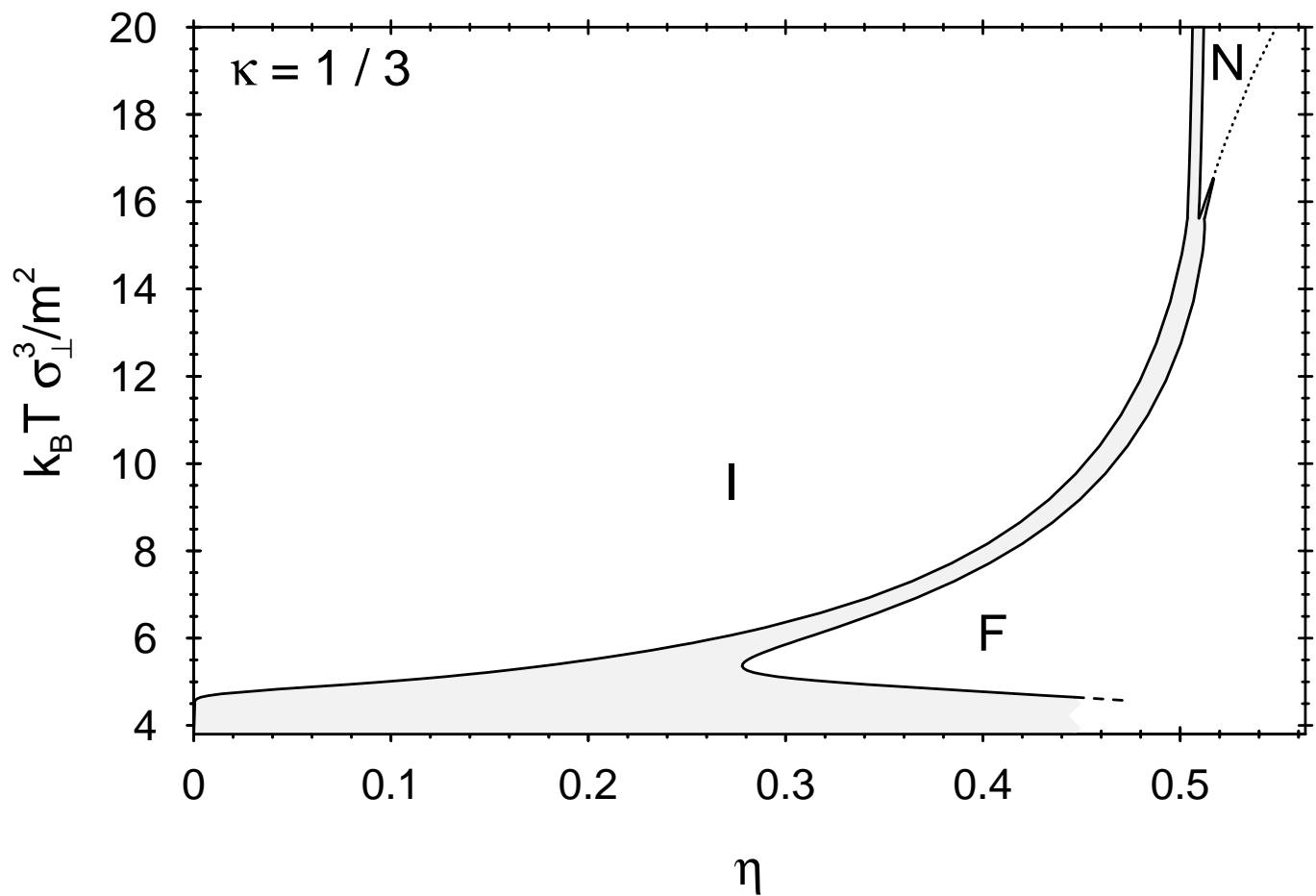
\begin{figure}[htbp]
\caption{Phase diagram for oblate ellipsoids with aspect ratio
$\kappa=1/3$. The I-N coexistence packing fractions for $T\to\infty$
are the same as those for $\kappa=3$, but in contrast to
\abbpref{fig:pdk3} a ferromagnetic phase is stable at high densities.}
\label{fig:pdk.33}
\end{figure}

The  properties of this model are determined by the packing fraction $\eta$,
the dimensionless ratio $k_B T \sigma_\perp^3/m^2$, where
$\sigma_\perp$ and $\sigma_\parallel$ are the particle diameters
measured perpendicular and parallel to the
dipole moment, respectively, and the aspect ratio
$\kappa=\sigma_\parallel/\sigma_\perp$ of the particles. We find that
for elongated particles the isotropic-nematic transition is shifted to
lower densities by the dipole moment. This trend is agreement with
earlier results from the Onsager virial theory \cite{Griechen},
integral equation theory \cite{PateyEll}, and a different version of
density-functional theory \cite{Vega}.
The two-phase region widens into gas-nematic coexistence at high dipole moments or
low temperatures (\abbref{fig:pdk3}). For ellipsoids of aspect ratio
$\kappa\lesssim 2$ (\abbref{fig:pdk2}) a ferromagnetic fluid (F)
appears at high densities, as well as a reentrant nematic phase at
high dipole moments that most probably is an artefact of the
theoretical approximations. The ferromagnetic phase is reached from
the nematic or isotropic states by continuous (dotted lines) or weakly
first-order (solid lines) transitions, depending on the temperature
and particle aspect ratio. Similar phase diagrams are found for oblate
particles ($\kappa<1$, see \abbref{fig:pdk.33}). If one compares fluids
with aspect ratios $\kappa$ and $1/\kappa$ the isotropic-nematic
coexistence packing fractions are the same for $m=0$ (or $T\to\infty$)
but the
ferromagnetic phase becomes stable at much higher values of $k_B T V_0/m^2$ ($V_0$ is the particle volume) for a given packing fraction
if $\kappa<1$, because in this case the dipole moments can approach
each other closer in the most favorable head-to-tail
configuration. For the same reason the ferromagnetic liquid should be
easier to attain in fluids of oblate than of spherical particles. For
dipolar hard spherocylinders one finds an analogous series of phase
diagrams as for elongated ellipsoids \cite{Diplc}.

\section{Gas-liquid interface} \label{lginterf}

Up to this point we were only concerned with bulk properties of dipolar
fluids and the influence of sample surfaces. If one allows for the
presence of interfaces between coexisting bulk phases various  new
phenomena  arise. By imposing appropriate boundary condition in
principle one can examine interfaces between any two phases that can coexist according to the phase diagrams shown
in the previous sections. These interfaces can be studied with the
same density-functional theories, although the calculations will be
much more involved due to the lower symmetry of
$\hat\rho(\rv,\omega)$. Here we restrain ourselves to the simplest
case, i.e., the interface between the isotropic gas and the isotropic liquid
\cite{TelodaGama,Frodl1,Frodl2}. For this system the number density  depends on the  coordinate
$z$ normal to the interface and the local orientational order near the interface is described by
the first nontrivial coefficient $\alpha_2(z)$ in an expansion in
terms of Legendre polynomials:
\begin{equation}
 \hat\rho(\rv,\omega)=\hat\rho(z,\cos\theta)=\frac{1}{2\pi}
 \rho(z)\left(\half +\alpha_2(z) P_2(\cos\theta)+\ldots\right).
\end{equation}

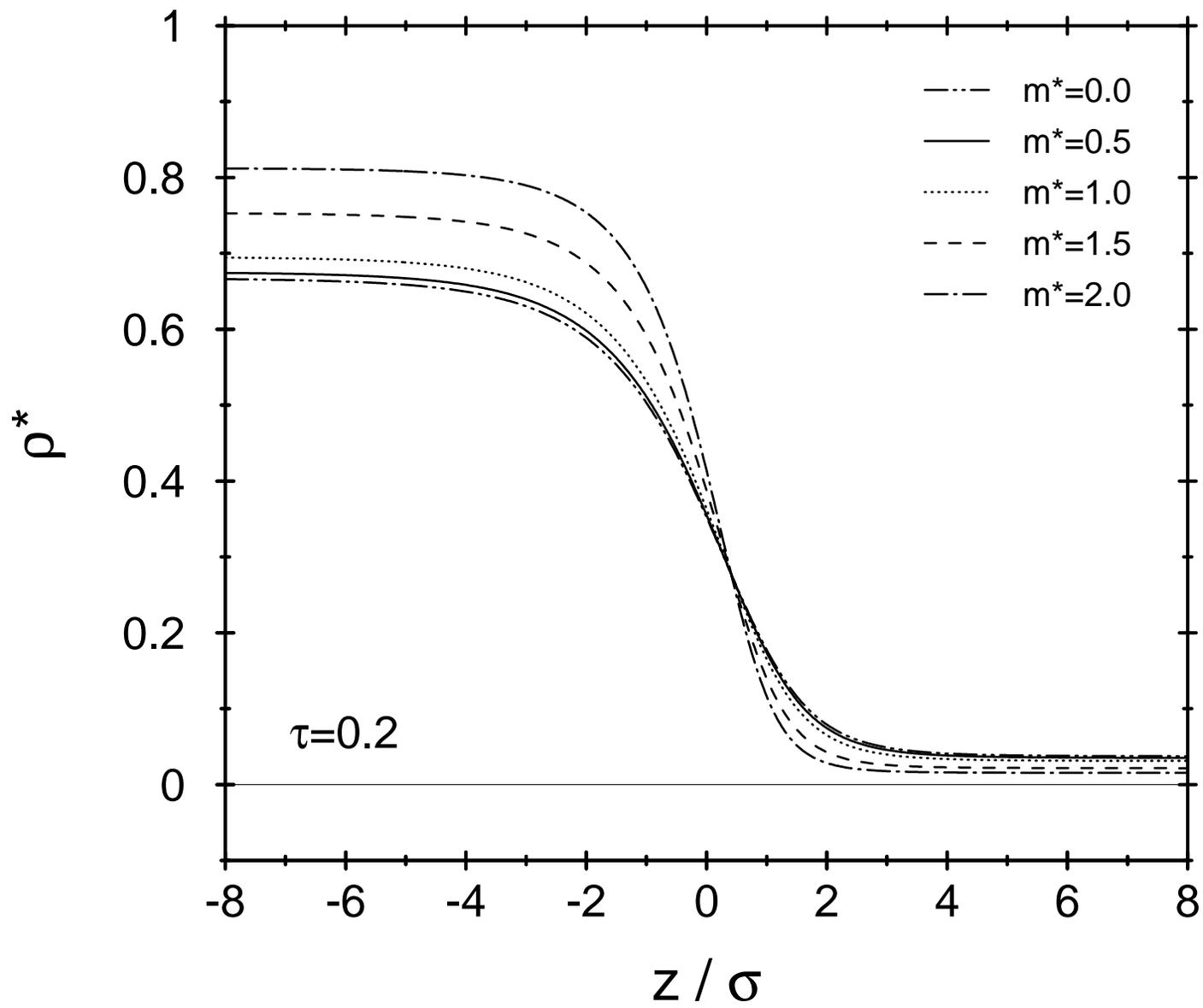
\begin{figure}[htbp]
\caption{Density profiles for the gas-liquid interface of a Stockmayer
fluid for various dipole moments at a fixed reduced temperature $\tau=1-T/T_c=0.2$.}
\label{fig:rhoz}
\end{figure}

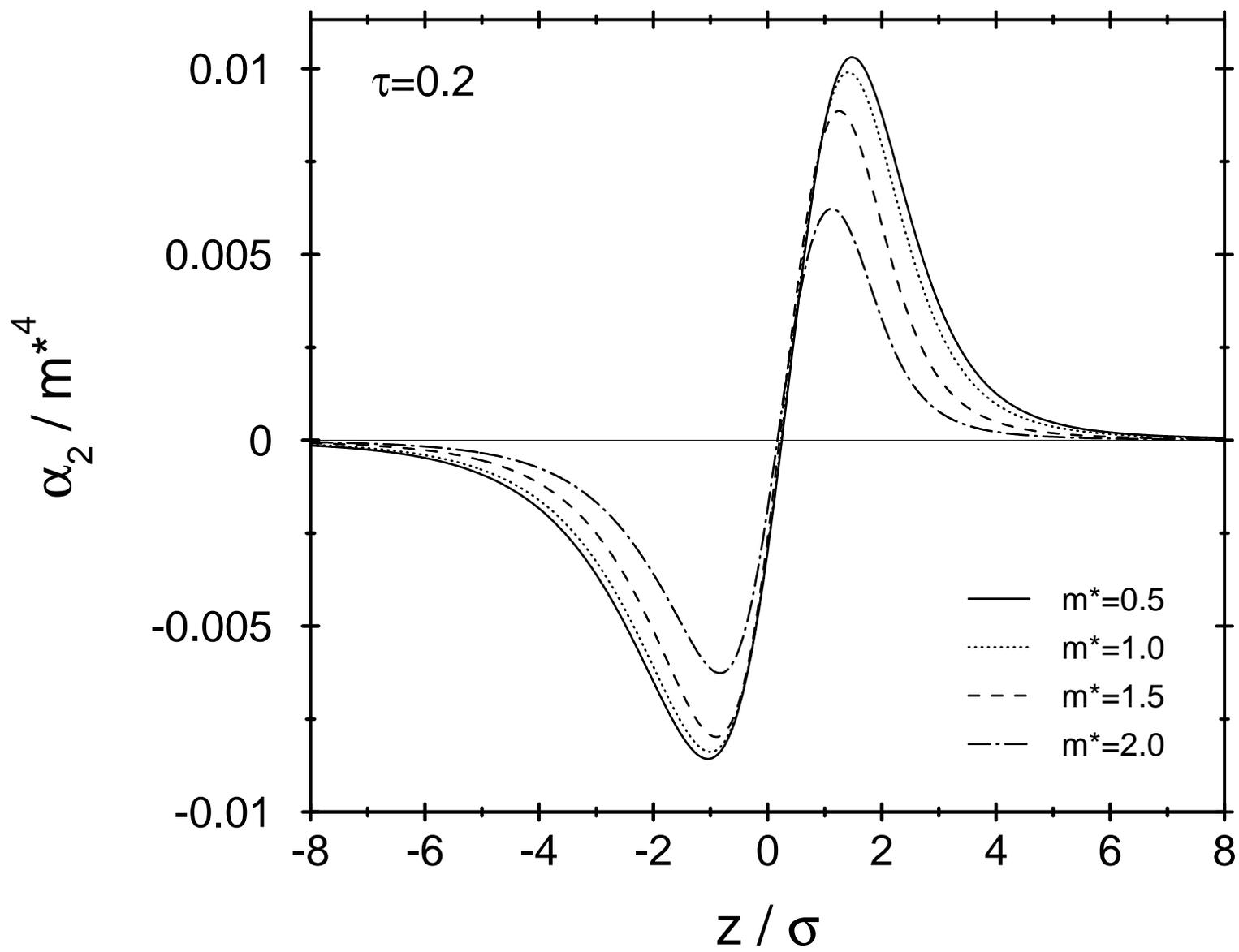
\begin{figure}[htbp]
\caption{Orientational order profiles $\alpha_2(z)$ at the gas-liquid
interface for the same parameters as in \abbpref{fig:rhoz}. Note that
$\alpha_2$ is scaled by $\ms^4$. The dipoles are aligned
preferentially parallel to the interface on the liquid side and perpendicular to the
interface on the gas side. For $\ms\to 0$ the profiles approach a
limiting form $\alpha_2(z/\sigma,\ms)=\ms^4 \alpha_2^{(0)}(z/\sigma)$.}
\label{fig:a2z}
\end{figure}

(The higher coefficients $\alpha_{l>2}(z)$ are very small.) The
functions $\rho(z)$ and $\alpha_2(z)$ have been calculated 
numerically by solving the integral
equation that results from the stationary condition for the density functional.
The  density profiles are monotonic and for small reduced dipole
moment $\ms$ they depend
only weekly  on  $\ms$ for fixed
distance $\tau=1-T/T_c$ from the critical temperature $T_c$ (see
\abbref{fig:rhoz}). Near the interface the particles are
preferentially aligned parallel to the interface ($\alpha_2(z)<0$) on
the liquid side and perpendicular to the interface ($\alpha_2(z)>0$) on the vapor
side (\abbref{fig:a2z}). The amplitude of $\alpha_2(z)$ increases
strongly ($\sim \ms^4$) with the dipole moment. Near the critical
point both the density profile 
\begin{equation}
  \rhos(z,\tau\to 0) =  \rhos_c+A_\rho(\ms) \tau^\beta
  F_\rho(z/\xi)\\
\end{equation}
and the orientational profile
\begin{equation} \label{a2profil}
  \alpha_2(z,\tau\to 0)  =  A_{\alpha_2}(\ms) \tau^{\beta+2\nu}
  F_{\alpha_2}(z/\xi)
\end{equation}
exhibit a scaling behavior.
Here $\xi$ is the bulk correlation length, $F_\rho$ and $F_{\alpha_2}$
are scaling functions, and $A_\rho$ and $A_{\alpha_2}$ are
nonuniversal amplitudes. Within the density functional approach the critical exponents $\beta$ and $\nu$ of
the order parameter and of the correlation length take on their mean-field
value $\half$. Equation~(\ref{a2profil}) has been used to interpret
ellipsometric measurements of the liquid-liquid interface of the
critical ionic mixture triethyl-n-hexylammonium triethyl-n-hexylboride
dissolved in diphenyl ether \cite{Law}.

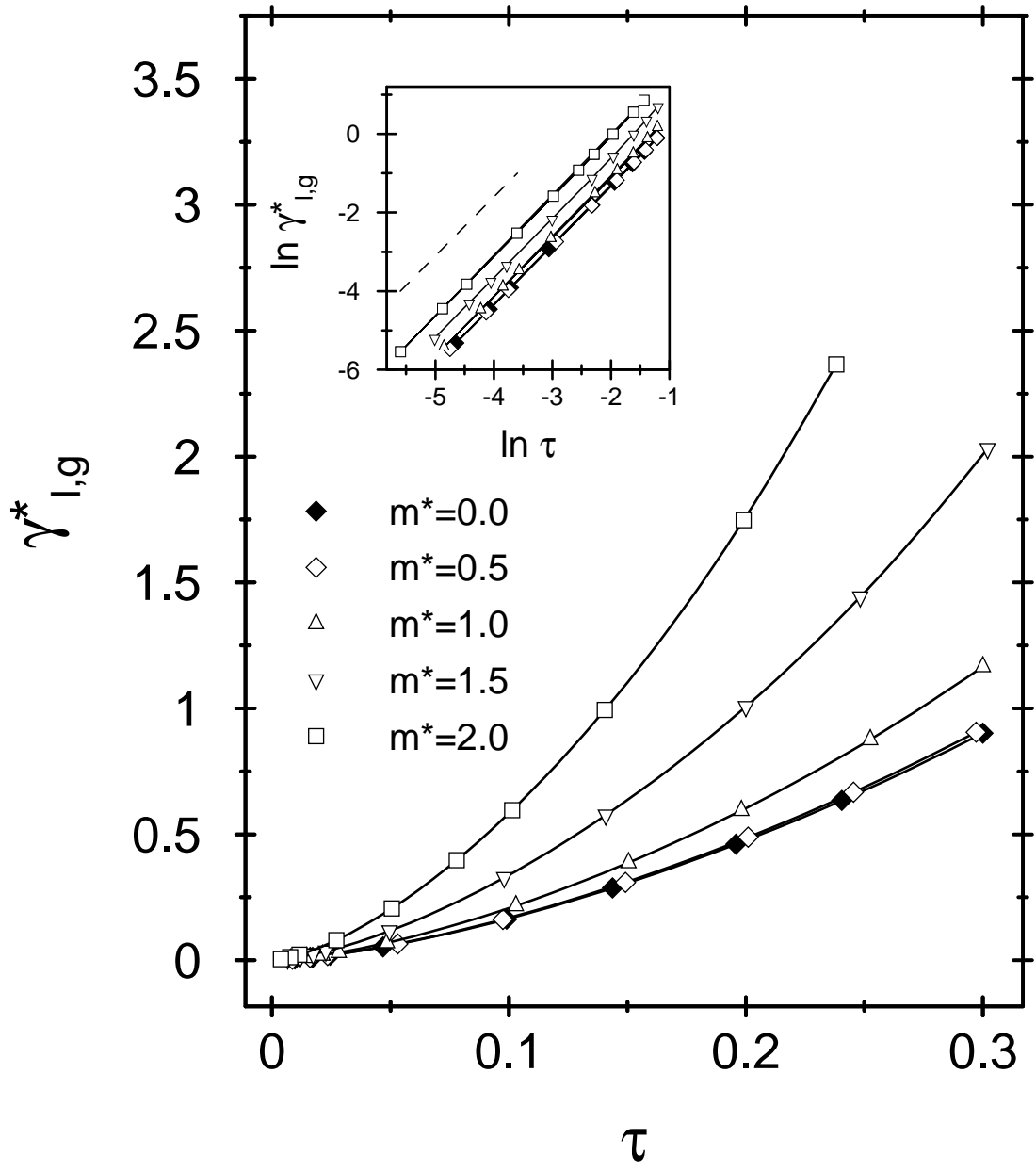
\begin{figure}[htbp]
\caption{Temperature dependence of the liquid-gas surface tension
$\gamma_{l,g}^\ast=\gamma_{l,g} \sigma^2/\epsilon$ for various dipole
moments. The double logarithmic plot in the inset demonstrates that
$\gamma_{l,g}\sim \tau^{3/2}$ for $\tau\to 0$.}
\label{fig:surft}
\end{figure}

The temperature dependence of the surface tension $\gamma$, which can
be calculated from these profiles, is shown in \abbref{fig:surft}. The
limiting power law $\gamma(\tau\to 0)=\gamma_0 \tau^{3/2}$ is valid even quite
far away from $T_c$. The amplitude $\gamma_0$ increases with
increasing dipole strength. The values of $\gamma$ are (for $\ms\leq
2$) practically unaltered if the orientational profile $\alpha_2(z)$
is replaced by $\alpha_2(z)\equiv 0$. This means that the main influence
of the dipolar interaction on the surface tension is due to the change
of the bulk phase diagram and of the number density profile but not due
to preferential orientations at the interface. This result has been
exploited for obtaining estimates of the surface tension of polar
fluids from other, more simple analytical approaches and simulations \cite{Abbas}.


\end{document}